\documentclass[12pt,final,1p,times]{elsarticle}

\usepackage{pifont}
\usepackage{natbib}
\usepackage{geometry}
\usepackage{graphicx}
\usepackage{txfonts}
\usepackage{subfigure}
\usepackage{wrapfig}
\usepackage{color}
\usepackage{booktabs}
\usepackage{caption}
\usepackage{amsmath}
\usepackage{amssymb}
\usepackage{mathabx}  
\usepackage{url}
\usepackage{colortbl}
\usepackage{multirow}
\usepackage{fnbreak}
\usepackage{helvet}
\usepackage{xspace}
\usepackage{tikz}
\PassOptionsToPackage{hyphens}{url}\usepackage[bookmarks=false]{hyperref}
\hypersetup{
    colorlinks,
    citecolor=black,
    filecolor=black,
    linkcolor=black,
    urlcolor=blue
}
\usepackage{cleveref}

\usepackage[font=bf]{caption}

\newcommand{\gomez}{G\'{o}mez}

\newcommand{\etal}{\textit{et al.}}
\newcommand{\todo}[1]{\iffalse #1 \fi}
\newcommand{\dt}[1]{$\delta t$ }

\newcommand{\autopas}{\textit{AutoPas}\xspace}
\newcommand{\ladds}{\textit{LADDS}\xspace}
\newcommand{\openmp}{\textit{OpenMP}\xspace}
\newcommand{\coolmuc}{\textit{CoolMUC2}\xspace}
\newcommand{\legend}{\textit{LEGEND}\xspace}
\newcommand{\accessed}{Accessed: 2022-03-01}

\graphicspath{{figs/}}

\definecolor{Gray}{gray}{0.7}

\begin{document}

\begin{frontmatter}
\begin{center}
\textbf{3\textsuperscript{rd} IAA Conference on Space Situational Awareness (ICSSA)} \\ \vspace{0.1in} \textbf{Madrid, Spain} \\ \vspace{0.4in} \textbf{IAA-ICSSA-22-0000} \\
\textbf{Deterministic Conjunction Tracking in Long-term Space Debris Simulations}
\end{center}


\author[fml1]{\textbf{Pablo \gomez}}
\author[fml2]{\textbf{Fabio Gratl}}
\author[fml2]{\textbf{Oliver B\"osing}}
\author[fml1]{\textbf{Dario Izzo}}


\address[fml1]{\normalsize European Space Agency, Advanced Concepts Team, Keplerlaan 1, 2201 AZ Noordwijk,  +31-71-565-6565, pablo.gomez@esa.int, dario.izzo@esa.int 
}
\address[fml2]{\normalsize Technical University Munich, Department of Informatics, Boltzmannstraße 3 85748 Garching Germany, +49-89-289-18-618, f.gratl@tum.de, oliver.boesing@tum.de 
}

\end{frontmatter}

\begin{flushleft}
\end{flushleft}

\section*{}
Numerical simulations are at the center of predicting the space debris environment of the upcoming decades. In light of debris generating events, such as continued anti-satellite weapon tests and planned mega-constellations, accurate predictions of the space debris environment are critical to ensure the long-term sustainability of critical satellite orbits.

Given the computational complexity of accurate long-term trajectory propagation for a large number of particles, numerical models usually rely on Monte-Carlo approaches for stochastic conjunction assessment. On the other hand, deterministic methods bear the promise of higher accuracy and can serve to validate stochastic approaches. However, they pose a substantial challenge to computational feasibility.

In this work, we present the architecture and proof-of-concept results for a numerical simulation capable of modeling the long-term debris evolution over decades with a deterministic conjunction tracking model. For the simulation, we developed an efficient propagator in modern C++ accounting for Earth’s gravitational anomalies, solar radiation pressure, and atmospheric drag. We utilized \autopas, a sophisticated particle container, which automatically selects the most efficient data structures and algorithms.

We present results from a simulation of 16 024 particles in low-Earth orbit over 20 years. Overall, conjunctions are tracked for predicted collisions and close encounters to allow a detailed study of both. We analyze the runtime and computational cost of the simulation in detail. In summary, the obtained results show that modern computational tools finally enable deterministic conjunction tracking and can serve to validate prior results and build higher-fidelity numerical simulations of the long-term debris environment.
\section{Introduction}

In recent decades the problem of space debris has become a central factor in mission design, spacecraft operations, and the question of the long-term sustainability of important orbits. Already, more than 5-10\% of total mission costs for low-Earth orbit (LEO) satellites are protective and mitigation costs~\cite{undseth2020space}. With the sharp increase in the number of launched payloads in recent years~\cite{lemmens2020esa}, the challenges caused by space debris are poised to increase further.

Consequently, modeling efforts of the space debris environment over a long time frame have become increasingly relevant in the last decades. The most well-known tool for long-term debris environment simulation is NASA's \legend software~\cite{liou2004legend} but there are other tools such as ESA's \textit{DELTA} software~\cite{virgili2016delta}.

A central challenge in long-term modeling lies in dealing with the need to simulate ever-growing numbers of particles (satellites and debris) with a fine temporal resolution for accurate propagation and conjunction tracking -- with timesteps on the scale of seconds or less -- while the overall simulation time spans decades, requiring millions if not billions of necessary iterations depending on the setup. Furthermore, the question of whether two particles collide during a timestep is, fundamentally, a pairwise interaction requiring exhaustive checks of distances between the particles' positions. These two factors have made the deterministic simulation of the debris environment over a long time frame prohibitively expensive in terms of computational cost. Therefore, previous long-term models, such as \legend~\cite{liou2004legend}, rely on stochastic analyses of the dynamics and occurring collisions.

In another field concerned with large-scale particle simulation, computational molecular dynamics, similar challenges have been dealt with. Here, particles are propagated by computing short-ranged pairwise potentials and solving Newton's equations of motion in every iteration. Due to the fast decay of the employed potentials with growing distance, efficient algorithms are used. They resolve the pairwise interaction, which is computationally prohibitive in $O(N^2)$, in $O(N)$ by introducing a spatial cutoff beyond which forces are too small to contribute in a significant way~\cite{rapaport2004art}. Long simulation times are common, needing millions of timesteps for the equilibration of systems, or sampling of evolving properties~\cite{heinen2016communication}. 

Based on these advances, ESA and TU Munich are currently exploring these tools in an ongoing collaboration studying the possibility of using them to model the long-term space debris environment. This work presents the first preliminary findings obtained from the newly created \textit{Large-scale Deterministic Debris Simulation} (\ladds) implemented in \texttt{C++17}. In particular, we study the viability of modeling objects in LEO, which given their speed are most challenging, over a period of 20 years. Detailed descriptions of \ladds together with an extensive analysis of the observed conjunctions and performance follow. All code for this project is available online under an open-source license.\footnote{\url{https://github.com/esa/LADDS/} \accessed}
\section{Methods}
From a high-level view, the simulation has two main computational steps: Propagating the simulated particles and checking whether conjunctions occur. This section will describe the employed algorithms and their integration into \ladds.

\subsection{Propagator}
\label{sec:propagator}

Overall, the implemented propagator aims to be computationally cheap while modeling all essential perturbations, which affect the particles' trajectories in LEO. Efficient parallelization is a critical factor for high performance. Hence, integrators which use adaptive timesteps are challenging to parallelize as individual particles don't share a timestep size. Thus, while higher accuracy integrators are available \cite{biscani2021revisiting} they were in their native form not suitable for this work. Instead, we implemented a fully parallel propagator relying on a fixed timestep. The modeled perturbations are based on Keplerian orbits combined with spherical harmonics (J2, S22, C22), solar radiation pressure, and atmospheric drag. Since we target an LEO population, lunar and solar gravitational forces are considered negligible. The concrete equations are similar to the formulations described by Vallado~\cite{vallado2001fundamentals}.

In summary, we use the following descriptions:
\begin{description}
    \item[Keplerian Motion]
    \begin{align}
        \Vec{a}_{Kep} = - \frac{G(m_{\Earth})}{|\Vec{x}|^3}\Vec{x}
    \end{align}
    with $\Vec{x}$ position relative to Earth core, $G$ the gravitation constant and $m_\Earth$ the mass of Earth.
    \item[Spherical Harmonics (inhomogeneous gravity field)]\hphantom{TheAlgorithm}\\
    The potential $U$ is given as
    \begin{align}
        U = \frac{Gm_\Earth}{|\Vec{x}|} \left(1 + \sum_{l=2}^\infty \sum_{m=0}^l \left(\frac{R_\Earth}{|\Vec{x}|}\right)^l P_{lm}\left(\sin\left(\phi_{gc}\right)\right) \left(C_{lm}\cos\left(m\lambda_{gc}\right) + S_{lm}\sin\left(m\lambda_{gc}\right) \right) \right)
    \end{align}
    with $R_\Earth$ as the radius of Earth, $\phi_{gc}$ the geocentric longitude, $P_{lm}$ the Legendre function, $\lambda_{gc}$ the object's polar coordinates, and $C_{lm}$ and $S_{lm}$ are empirical coefficients determined from observation data of satellites orbiting the Earth~\cite{long1989goddard}.
    The acceleration follows from the partial derivatives of the potential
    \begin{align}
        \Vec{a}_{harmonics} = \frac{\partial U}{\partial | \Vec{x}|} \left(\frac{\partial |\Vec{x}|}{\partial \Vec{x}}\right)^T + \frac{\partial U}{\partial \phi_{gc}} \left(\frac{\partial \phi_{gc}}{\partial \Vec{x}} \right)^T + \frac{\partial U}{\partial \lambda_{gc}} \left(\frac{\partial \lambda_{gc}}{\partial\Vec{x}}\right)^T.
    \end{align}
    Note that this formulation also includes the zonal terms.
    \item[Solar Radiation Pressure]
    \begin{align}
        \Vec{a}_{SRP} = \texttt{AU}^2 P_{SRP}\frac{A}{m}\frac{\Vec{x} - \Vec{x}_\Sun}{|\Vec{x} - \Vec{x}_\Sun|^3}
    \end{align}
    with \texttt{AU} as the astronomical unit, $P_{SRP} = 4.56 \cdot 10^-6 \dfrac{N}{m^2}$ the solar radiation pressure at 1 \texttt{AU}, $A$ object's area facing the sun, $m$ the object's mass, and $\Vec{x} - \Vec{x}_\Sun$ the distance between the object's and Sun's center.
    \item[Atmospheric Drag]
    \begin{align}
        \Vec{a}_{Drag} = - \frac{C_d A}{2 m} p(h) |\Vec{v}_{rel}|^2 \frac{\Vec{v}_{rel}}{|\Vec{v}_{rel}|}
    \end{align}
    with $C_d$ as the drag coefficient, $A$ the object's (estimated) cross section area, $m$ the object's mass, and $p(h)$ being the atmospheric density at the object's altitude $h$ above ground based on~\cite{picone2002nrlmsise}. $\Vec{v}_{rel}$ is the object's velocity relative to Earth's atmosphere.
\end{description}

\subsection{Conjunction Tracking}
\label{sec:conjunctionTracking}
The conjunction detection method is the second critical component for simulating the long-term occurrence of conjunctions. Given that this work aims to showcase the potential of a deterministic simulation, previous approaches with probabilistic collision estimates, such as the \textit{Cube} approach \cite{liou2003new,lewis2012synergy}, are not applicable. Instead, we propose a method -- similar to Alarcon \etal{} \cite{alarcon2002collision} -- that is based on the deterministic positions, size and operational status of the particles. To keep computational costs manageable, we assume spherical particle shapes for the purposes of conjunction detection. Furthermore, we use a sub-timestep interpolation to achieve a higher temporal resolution than the timestep \dt{} used for the propagation.

\begin{figure*}[t!]
    \begin{center}
        \includegraphics[width=0.75\textwidth]{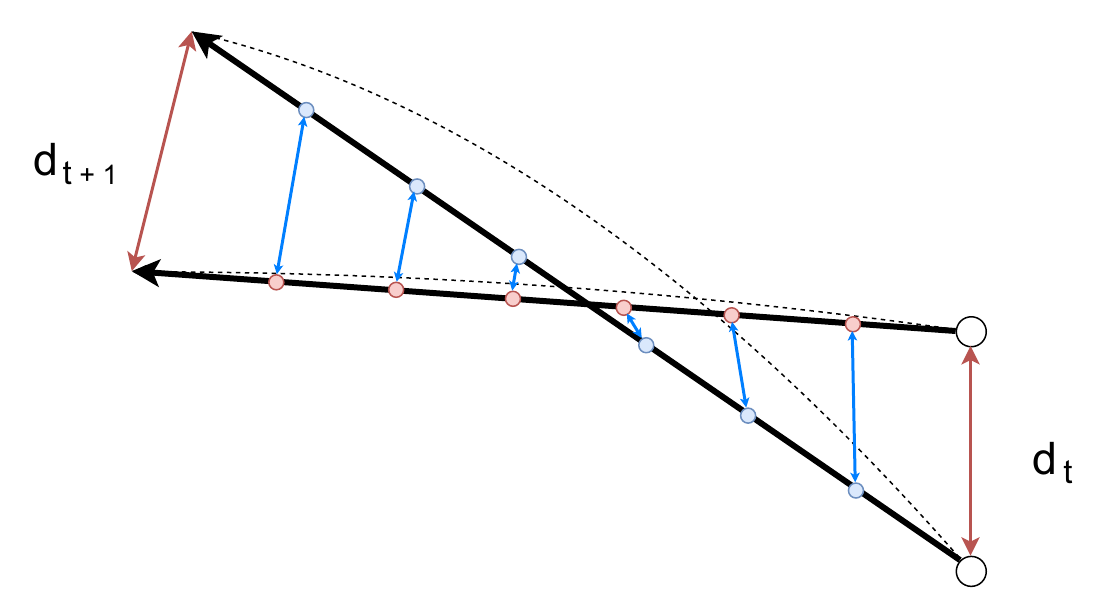}
        \caption{\small{Schematic illustrating the actual trajectories of two particles (dashed lines), linear interpolations (thick lines) with particle positions and distances at some points in time (as dots and blue lines) as well as the naive closest encounters at $\tau_t$ and $\tau_{t+1}$ (red lines). The interpolation allows a sub-timestep estimation of the distance at closest encounter $d$ that is more accurate.}}
        \label{fig:interpolation}
    \end{center}
\end{figure*}

More precisely, given two particles $p_1$ and $p_2$, their locations $x^t_1$,$x^t_2$ and velocities $v^t_1$,$v^t_2$ at timestep $t+1$, and radii $r_1$,$r_2$, we compute their closest approach $d$ in the interval $[\tau_{t},\tau_{t+1}]$ using a linear interpolation as depicted in Figure \ref{fig:interpolation}. This allows an analytic formulation that is more precise than using $d_t$ or $d_{t+1}$ while remaining computationally cheap. Thus, the time of the closest encounter during this timestep $\tau_d$ is given by

\begin{equation}
    \tau_d = \frac{x^t_1 v^t_1 - x^t_1 v^t_2 + x^t_2 v^t_2 - x^t_2 v^t_1}{v^t_1 v^t_1 + v^t_2 v^t_2 - 2 v^t_1 v^t_2}
\end{equation}

and consequently, the closest encounter distance is given by

\begin{equation}
    d =  \begin{cases}
        ||x^t_1 - x^t_2||_2 & \text{for } \tau_d < 0\\
        ||x^t_1 + \tau_d v^t_1 - x^t_2 - \tau_d v^t_2||_2 & \text{for } 0\leq \tau_d\leq \delta t\\
        ||x^t_1 + \delta t v^t_1 - x^t_2 - \delta t v^t_2||_2 & \text{for } \delta t < \tau_d
        \end{cases} .
\end{equation}
Based on this formulation, we define a collision as an encounter where $d \leq (r_1+r_2)$. In light of unavoidable numerical errors and to study the impact the estimated particle radii, we introduce the additional parameter $\kappa$ to study all conjunctions below some distance. Thus, a conjunction occurs when $d \leq \kappa (r_1+r_2) $. We tracked conjunctions for $0 < \kappa \leq 10$.

Furthermore, the simulation distinguishes between passive objects and actively operated satellites as defined by the \textit{CelesTrak SATCAT Operational Status}\footnote{\url{https://celestrak.com/satcat/status.php} \accessed} codes. Conjunctions between actively operated satellites are discarded as these would presumably be evaded in daily satellite operations. Conjunctions between active satellites and passive objects are only considered if the passive object has a radius smaller than 10~cm to model the challenges in tracking small objects \cite{gomez2017description}. Finally, as this initial work focuses only on conjunctions and not simulating resulting breakups, satellites remain active after a conjunction. To account for this, only the closest encounter two objects had throughout the simulation is considered to avoid repeated encounters during consecutive timesteps.

\subsection{Implementation}
The implementation of \ladds follows a modular design, compartmentalizing any state within each component. This method increases the comprehensibility of the code and facilitates exchanging components to explore the impact of, e.g., different propagation or conjunction tracking algorithms.

The first of the major components is the storage for all simulated particles. For this, we employ \autopas,\footnote{\url{https://github.com/AutoPas/AutoPas} \accessed} a \texttt{C++17} node-level high-performance library for arbitrary, short-range N-Body simulations. Here, it acts as a black-box particle container, which means that the simulation is oblivious to the actual data structure used within \autopas. This allows \autopas to freely choose its internal data structures and algorithms, such as Linked Cells or Verlet Lists, as well as algorithmic optimizations, e.g. exploiting symmetries in potentials to reduce the number of computations. For more information about the possible configurations refer to the \autopas release paper \cite{gratl2021nways} or the official documentation.\footnote{\url{https://autopas.github.io/} \accessed} The library can even adapt this configuration at runtime multiple times over the course of the simulation -- a feature called dynamic automatic algorithm selection.

Designed for N-Body simulations, \autopas provides an interface for the efficient pairwise interaction of spatially close particles. We use this feature for our conjunction tracking by implementing the algorithm presented in \Cref{sec:conjunctionTracking} within a functor which we pass to \autopas. During such a pairwise iteration step, \autopas then applies the functor to at least all particles within a given cutoff distance. If they are indeed within that cutoff and at least one particle is passive and has a radius smaller than 10~cm, we compute the linear interpolation to estimate the closest sub-timestep approach and whether a conjunction has taken place.

When compiled with \openmp,\footnote{\url{https://www.openmp.org/} \accessed} \autopas will automatically execute this pairwise interaction using shared memory parallelization, choosing the fastest algorithm while exploiting interaction symmetry and avoiding data races.

The second major component is the numerical propagator, responsible for simulating the motion of particles over time. For this we developed \textit{OrbitPropagator},\footnote{\url{https://github.com/FG-TUM/OrbitPropagator} \accessed} which implements a fourth-order Yoshida integrator \cite{yoshida1990construction} and the perturbations described in \Cref{sec:propagator}. The integrator was validated against \textit{heyoka} \cite{biscani2021revisiting}. For every position update, all particles stored in \autopas are piped through the propagator. Since all of these updates are independent of each other, this is easily parallelized in \openmp.

\begin{figure}
    \centering
    \includegraphics[width=\textwidth]{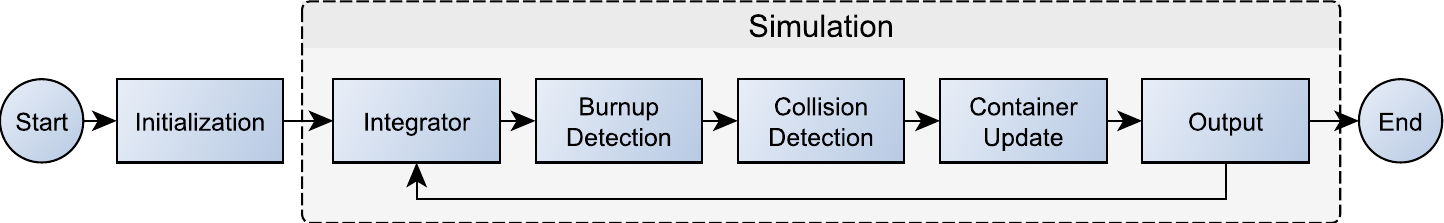}
    \caption{Flowchart of \ladds}
    \label{fig:flowchart}
\end{figure}

\Cref{fig:flowchart} shows the general flow of the simulation. After the initialization step, the rest of the simulation consists only of the main simulation loop, primarily alternating between the integrator and the conjunction detection. The frequency in which output is written can be configured by the user.

\section{Results}

\subsection{Dataset}
\label{sec:dataset}
The dataset for the presented results originates from the data provided by \textit{CelesTrak}\footnote{\url{https://celestrak.com/} \accessed} and the \textit{18th Space Control Squadron} on \textit{space-track.org}.\footnote{\url{https://www.space-track.org/} \accessed} Data from a total of $16\,024$ objects was used after filtering based on several criteria. The International Space Station (ISS) was removed from the data given its exceptional size (which makes collisions much more likely). Already operational satellites from the constellations by OneWeb and SpaceX (Starlink) were removed as these are being studied separately in a parallel study. Further, this work focused in particular on LEO where conjunctions are -- given the smaller volume -- proportionally likelier, thus only satellites operating between $175$ and $2000$ km above the ground were investigated.

\begin{figure*}[t!]
    \begin{center}
        \includegraphics[width=0.95\textwidth]{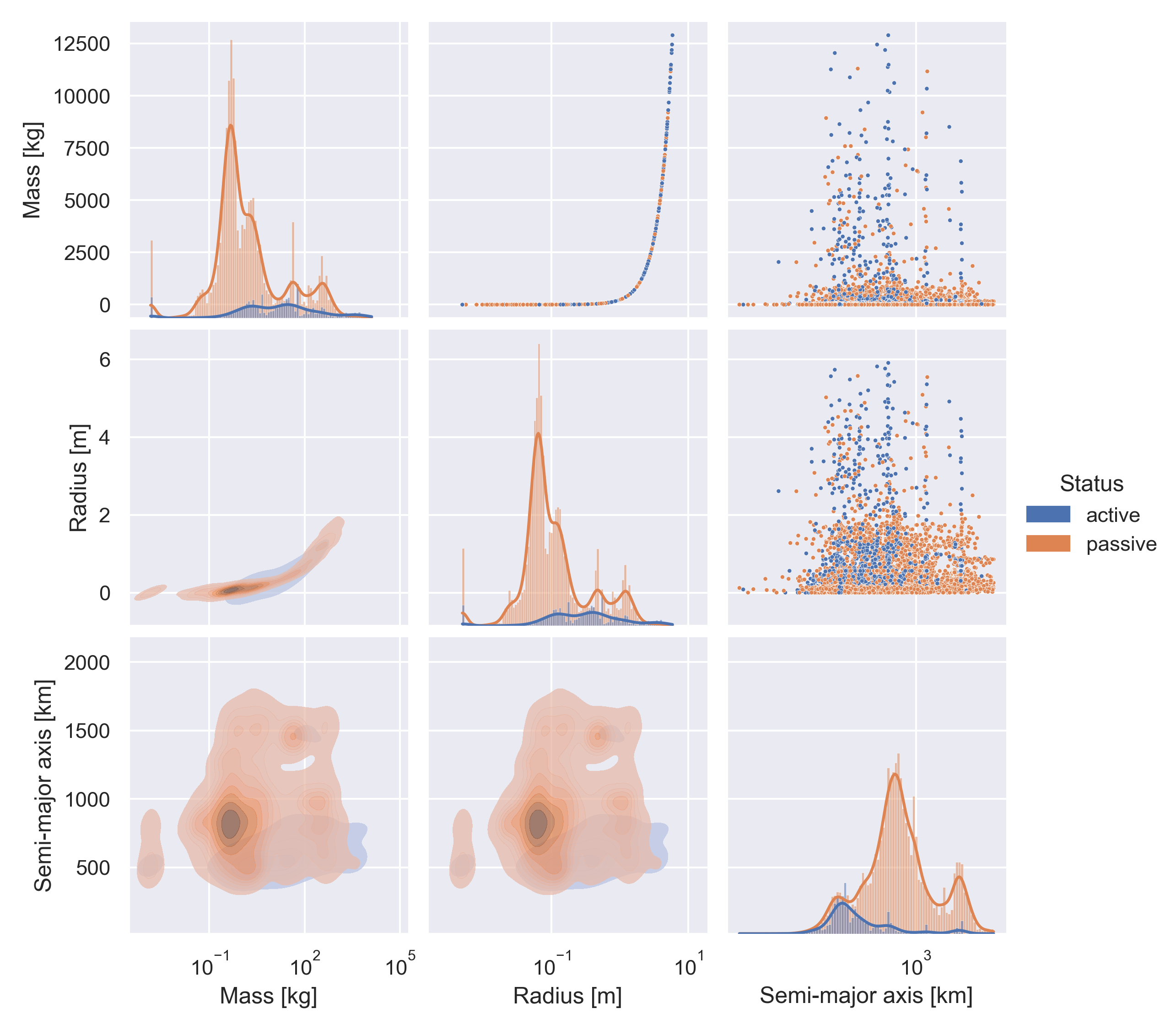}
        \caption{\small{Detailed distributions for active and passive objects in regard to mass, radius and semi-major axis of their orbit. Diagonal plots display histograms (with counts on the vertical axis), upper triangle scatters and the lower triangle kernel density estimates of the joint distributions.}}
        \label{fig:dataset}
    \end{center}
\end{figure*}

Only conjunctions between active and passive objects smaller than 10~cm radius or passive and passive objects are considered, respectively (see \Cref{sec:conjunctionTracking} ). Thus, based on the \textit{CelesTrak SATCAT Operational Status} the dataset contains $1996$ active and $14\,028$ passive objects. Similar to the work by Johnson \etal{} \cite{johnson2001nasa}, radii of objects were estimated as 
\begin{equation*}
r [m] = \sqrt{\frac{RCS}{\pi}}    
\end{equation*}
based on the radar cross section ($RCS$) and masses as 
\begin{align*}
m [kg] =  \begin{cases}
         \frac{4}{3} \, \pi \,r^3 \, 2698.9 \frac{kg}{m^3}& \text{for } r \leq 0.01m \\
         \frac{4}{3} \, \pi \, r^3 \,92.937 \frac{kg}{m^2} \, (2r)^{-0.74} & \text{for } r > 0.01m \\
    \end{cases} .
\end{align*}
These values were also used to compute the area-over-mass for the solar radiation pressure and atmospheric drag. If available on \textit{CelesTrak} the numerical $RCS$ from there was used. Otherwise, the \textit{space-track.org} $RCS$ classification into \textit{small} ($RCS < 0.1~m^2$), \textit{medium} ($0.1~m^2 < RCS < 1~m^2$) and \textit{large} ($1~m^2 < RCS$) was used to sample from normal distributions for these respective classes matching the data from \textit{CelesTrak} in terms of mean and standard deviation. No objects below $r=0.005~m$ were included. All objects were propagated to the time 2022-01-01 00:00:00.000 using the \textit{SGP4} propagator \cite{hoots2004history} implementation available through the Python module \textit{pykep} \cite{izzo2012pygmo}. BSTAR values used in the drag formulation (see \Cref{sec:propagator}) relied on the values provided in the two-line element set if available. Otherwise, they were approximated using above computed radius, mass and a drag coefficient $C_d$ of 2.2 \cite{cook1965satellite}. A detailed overview of the dataset's distribution of semi-major axis, mass and radius in relation to status is given in Figure \ref{fig:dataset}. As expected, most active objects are larger with more mass and many of them are in LEO. Passive objects are on average smaller, lighter, and have a higher semi-major axis.

\subsection{Simulation Setup}
All presented results were obtained on the \coolmuc\footnote{\url{https://doku.lrz.de/display/PUBLIC/CoolMUC-2} \accessed} segment of the \textit{Linux Cluster} at the \textit{Leibniz Supercomputing Centre}.\footnote{\url{https://www.lrz.de/} \accessed} As we only make use of shared memory parallelization the simulations were executed with 28 threads, which is a full node without hyper-threading. The code was compiled with \textit{GCC} version 10.2.\footnote{\url{https://gcc.gnu.org/} \accessed}

The actual simulations were conducted with a timestep $\delta t=10$~seconds. Particles are removed as soon as they reach an altitude of less than 150~km above ground as they are considered to be burning up. \autopas is restricted to finite, box-shaped domains. Therefore, the simulation is confined to a cube with a side length of 20000~km and its center at the center of the Earth. This size was chosen so that no particles will escape over the course of the simulation, so no boundary conditions need consideration.

\subsection{Observed Conjunctions}

\begin{figure}
\centering
{\includegraphics[width=0.6\textwidth]{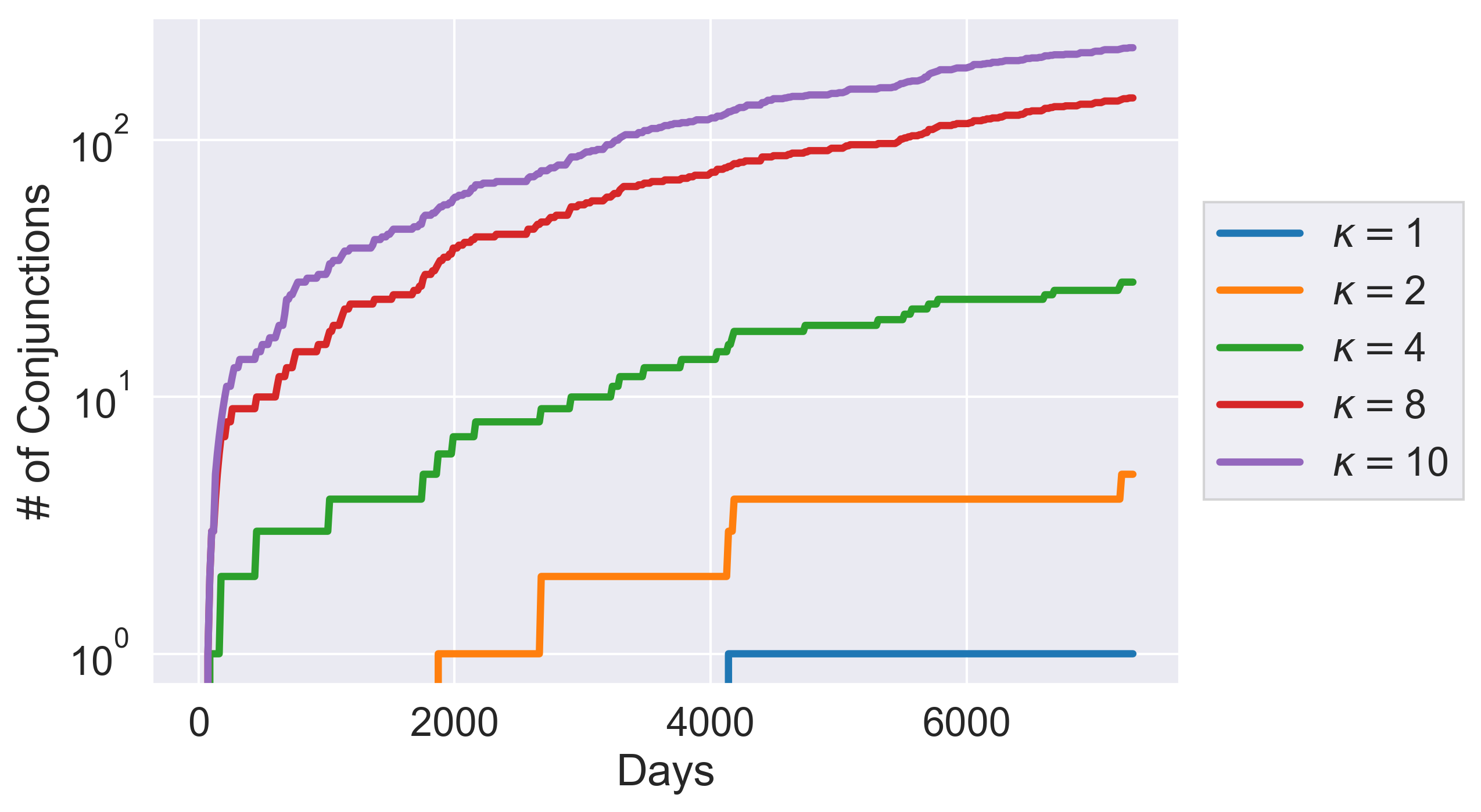}}
\hfill
{\includegraphics[width=0.34\textwidth]{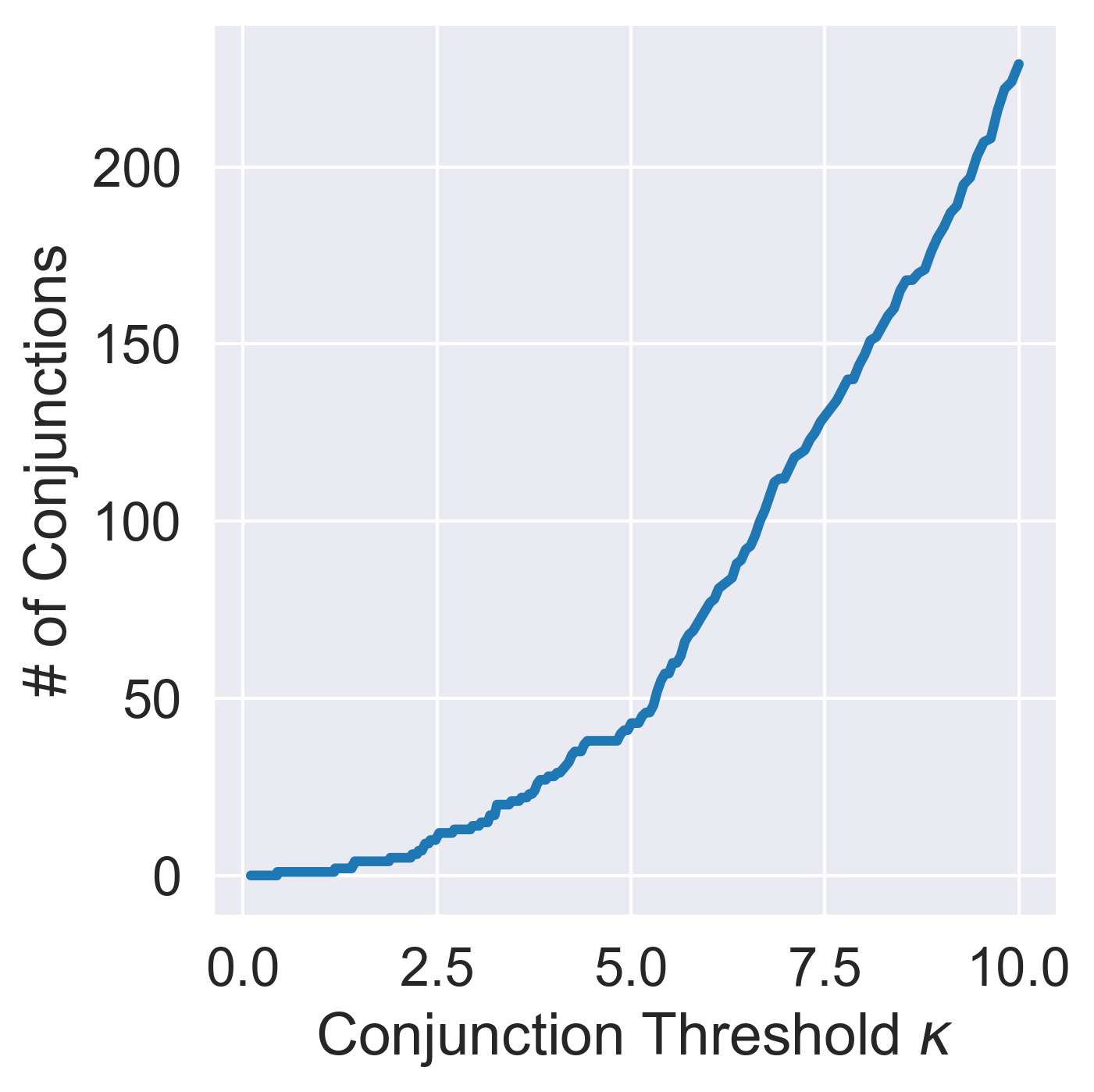}}
\caption{(left) Evolution of the number of conjunctions over time depending on the threshold $\kappa$;  (right) Detailed relationship between the total number of conjunctions and $\kappa$}
\label{fig:conj}
\end{figure}

In terms of conjunctions, several factors are noteworthy regarding the observations. First off, \Cref{fig:conj} (left) displays the observed conjunctions depending on $\kappa$ over the course of the simulation. Given the cubical impact of $\kappa$, the number of observed conjunctions rises by two orders of magnitude. Between 1 and 229 conjunctions are detected for $\kappa=1$ and $\kappa=10$, respectively. In total, 70 conjunctions happened between active and passive particles while 159 conjunctions involved only passive particles (active particles evade each other). \Cref{fig:conj} (right) displays the detailed relationship between the total number of observed conjunctions and $\kappa$. Overall, it is noteworthy that even though only one ($\kappa=1$) collision was found, the much more frequent conjunctions are also a relevant quantity as they may trigger collision avoidance maneuvers due to higher collision probabilities as monitored in spacecraft operations \cite{bacon20202019,merz2017current}.

\begin{figure}
\centering
{\includegraphics[width=0.45\textwidth]{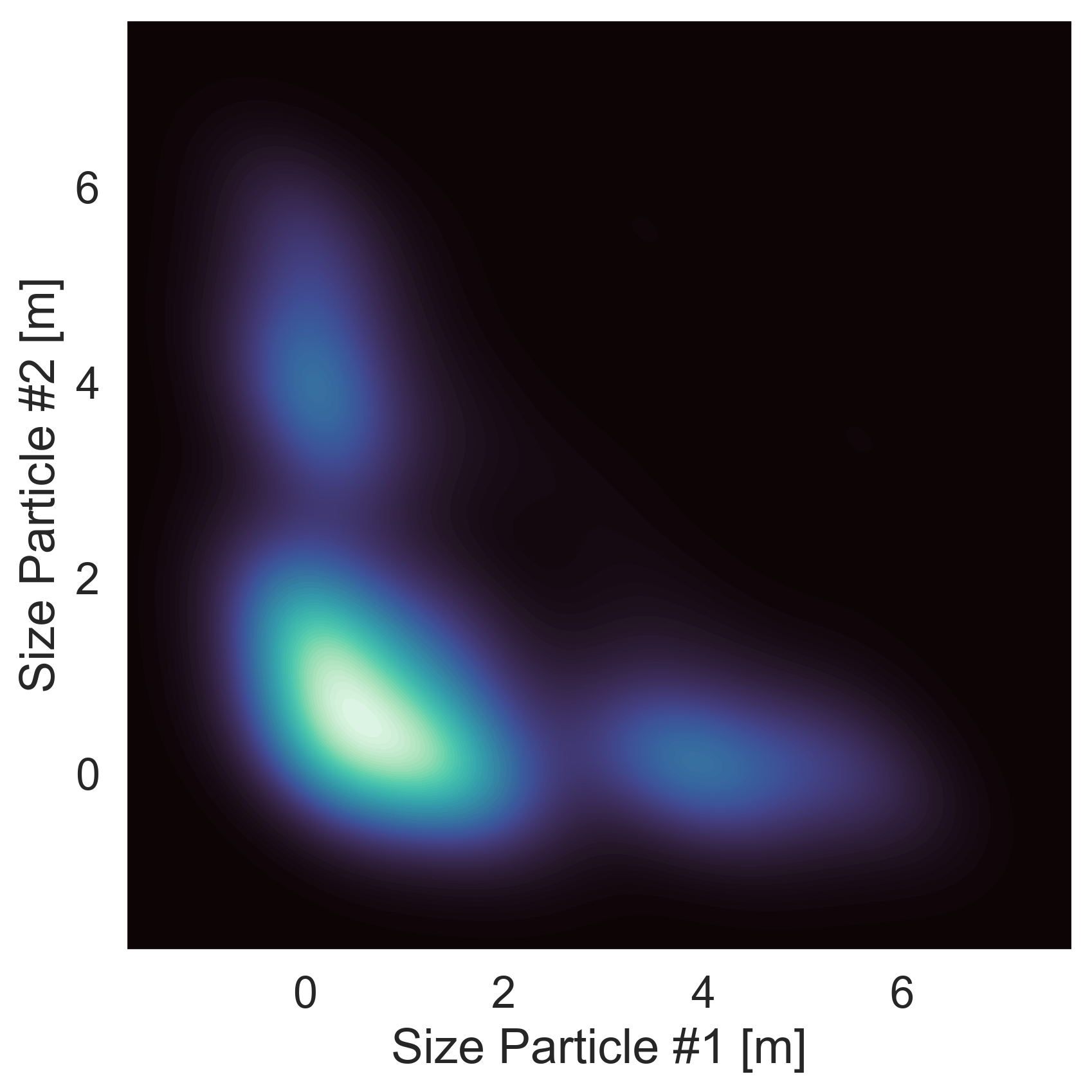}}
\hfill
{\includegraphics[width=0.45\textwidth]{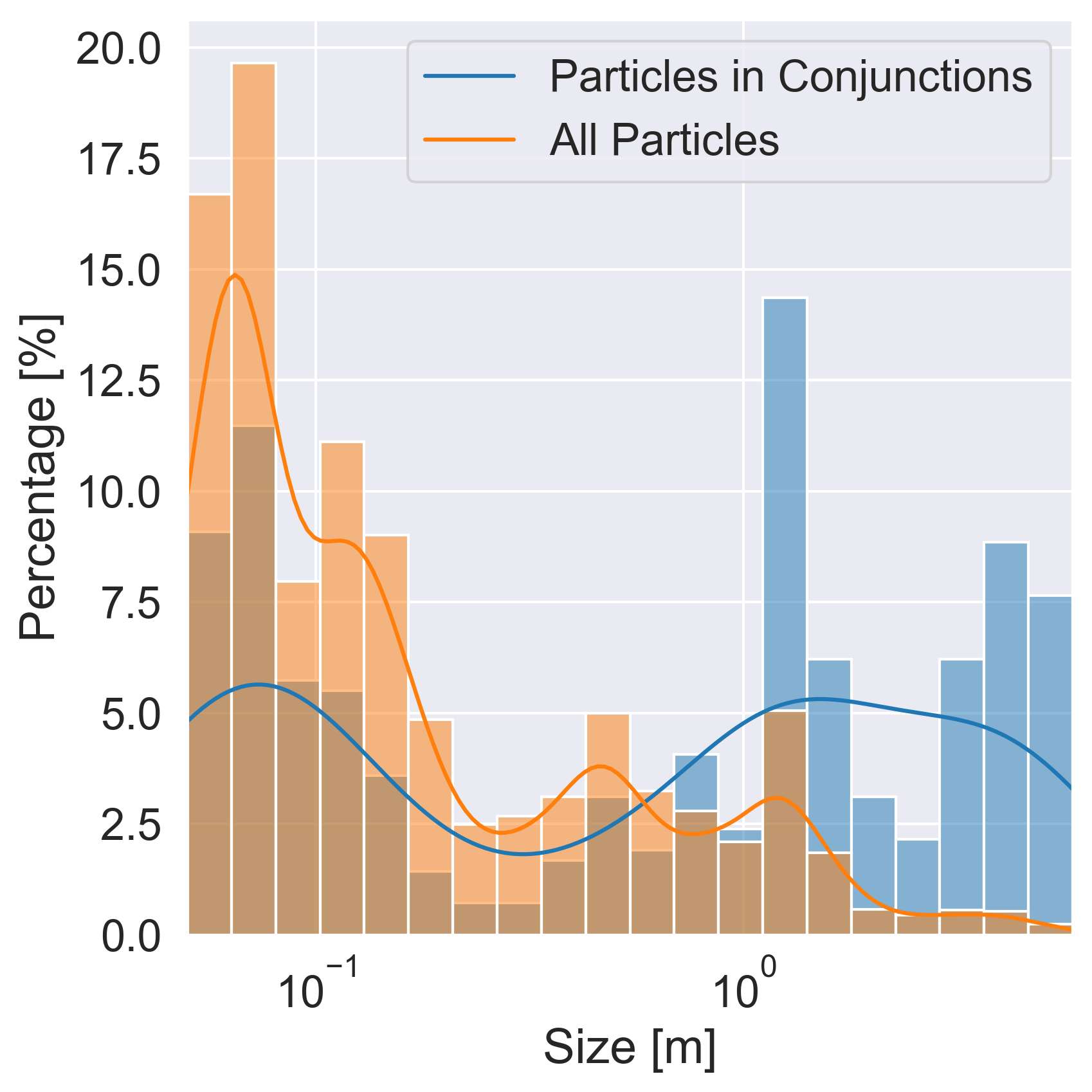}}
\caption{(left) Visualization of the sizes of objects in conjunctions with each other (commutative) with brightness indicating density; (right) Size of the particle population vs. the objects involved in conjunctions}
\label{fig:sizes}
\end{figure}

A second relationship worth monitoring is the one between size and conjunctions. As seen in \Cref{fig:sizes} (left), most conjunctions occur between large and small, and large and large particles. This is due to the higher likelihood of conjunctions given the objects' larger radii. The -- compared to large-with-small conjunctions -- small number of conjunctions between large and large particles is due to their overall lower occurrence rates and the higher ratio of active satellites with higher radii. \Cref{fig:sizes} (right) also confirms this as the size distribution of particles involved in conjunctions is clearly distinct from the overall population featuring a higher rate of large satellites. Note that collisions between large debris fragments are particularly critical as they will likely lead to a large number of new debris fragments. However, the very small particles below 10~cm radius also feature prominently in conjunctions as they are not evaded given the difficulty in tracking them (see \Cref{sec:conjunctionTracking}).

\begin{figure}
\centering
{\includegraphics[width=0.45\textwidth]{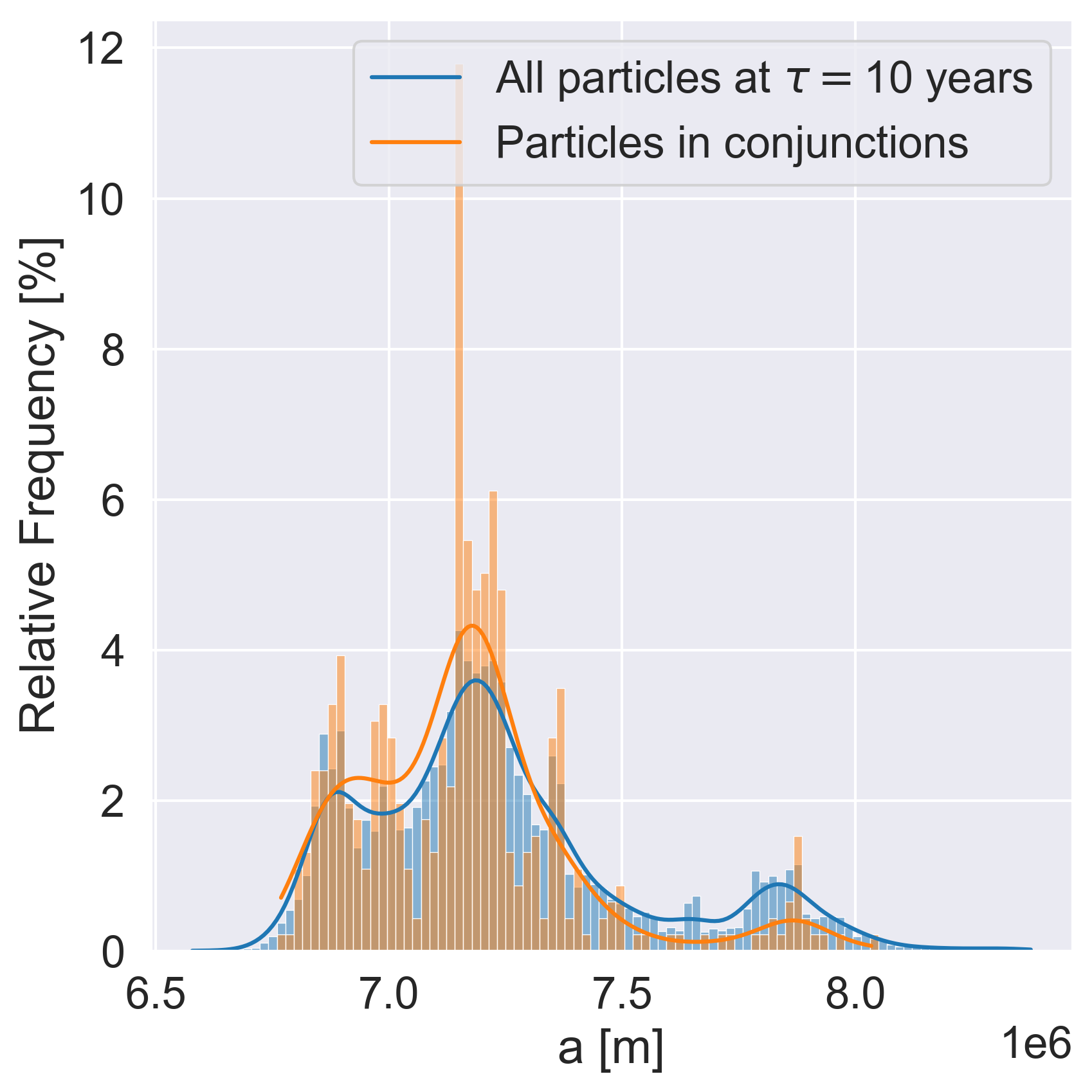}}
\hfill
{\includegraphics[width=0.45\textwidth]{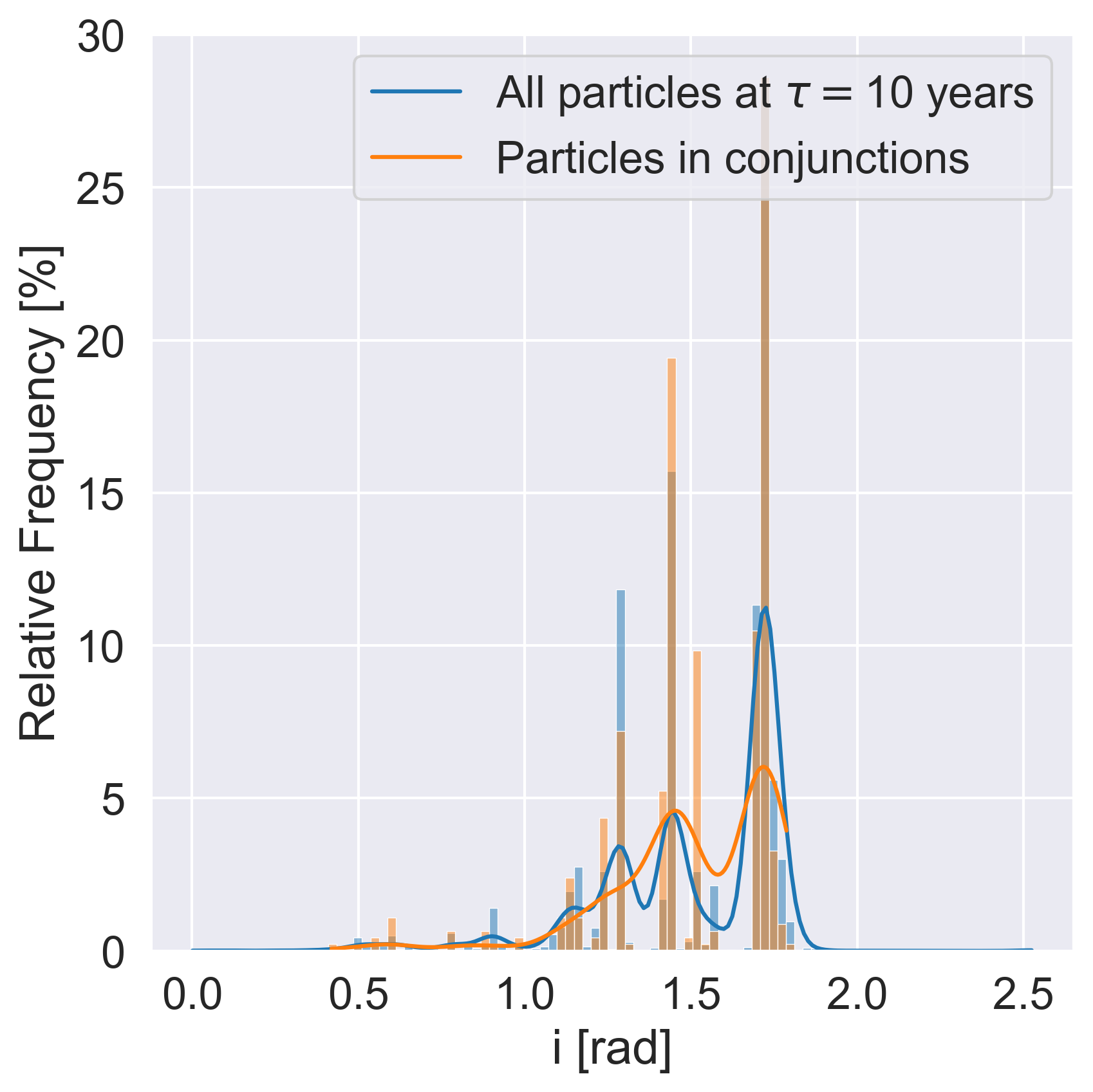}}
\caption{(left) Semi-major axes of the particle population vs. the objects involved in conjunctions; (right) Inclination of the particle population vs. the objects involved in conjunctions}
\label{fig:elements}
\end{figure}

Finally, a closer inspection of the orbital elements of objects in conjunctions is warranted. \Cref{fig:elements} (left) demonstrates that objects involved in conjunctions have a proportionally smaller semi-major axis compared to the overall population. This is particularly noteworthy since -- as explored in \Cref{sec:dataset} -- especially active satellites in the population have a smaller semi-major axis, whereas most of the passive objects are at higher semi-major axes. The other orbital elements do not show similarly clear relationships. Figure \ref{fig:elements} (right) displays the inclination of the orbits in conjunctions. There are some notable differences in the distributions, which feature different peaks, but the relationship is less clear compared to the semi-major axis.

\subsection{Performance Analysis}
\begin{figure}
    \centering
    \includegraphics[width=.7\textwidth]{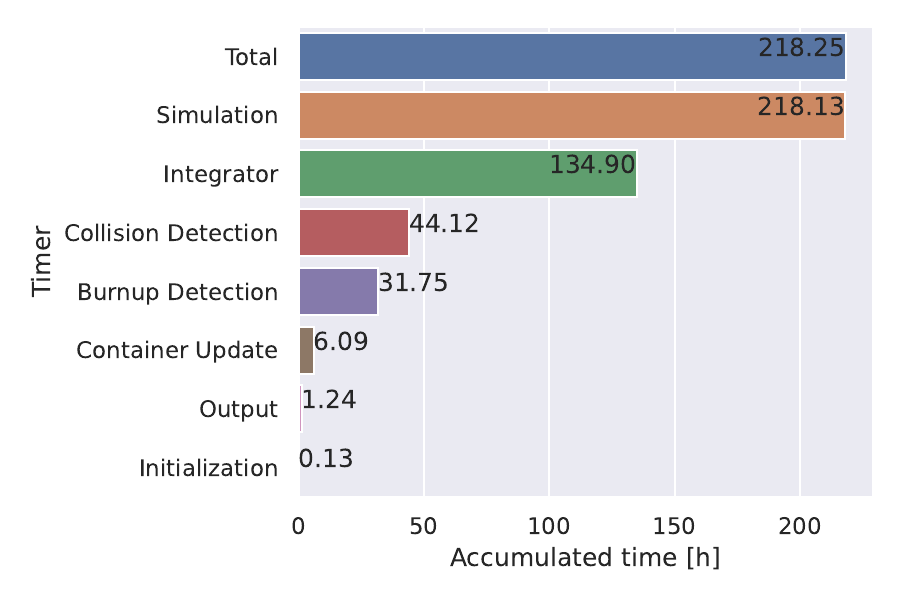}
    \caption{Accumulated runtime of steps over 20 years of the simulation. Labels correspond to the steps shown in \Cref{fig:flowchart}}
    \label{fig:runtime}
\end{figure}

In the following, a closer look into the overall performance and performance-critical elements of the simulation is taken.
To get an idea of which parts of the simulation are the most expensive, \Cref{fig:runtime} provides a breakdown of all major steps.

Due to timeout restrictions on the utilized cluster computer, the simulation was conducted in four runs, by saving and reloading checkpoints. The \texttt{Initialization} timer covers the loading of the checkpoints. It shows an insignificant share of the total time at less than 0.1\%. \texttt{Output} represents the time spent for writing the HDF5 files,\footnote{\url{https://www.hdfgroup.org/solutions/hdf5/} \accessed} which track all conjunctions as well as all particle positions every 1000 iterations. Thereby, we end up with over 63\,710 saved simulation states in a file of about 32~GB. The time spent on this is 0.5\% of the total time and thus acceptable given the file size. Also, the \texttt{Container Update}, which tracks the time for \autopas' bookkeeping to maintain the Linked Cells data structure, is reasonably quick with less than 3\%. \texttt{Burnup Detection} takes longer than initially expected with almost 15\% of the total time. This is due to the missing parallelization of this simulation step at the time of the long simulation. Later experiments with the same dataset and hardware showed that parallelizing the burn-up handling yields a speedup of 13 for this step. To put this into perspective, before the optimization \texttt{Burnup Detection} takes more than five times longer than the \texttt{Container Update}. After implementing the parallelization, it is more than twice as fast.
Interestingly, \ladds spends only about 20\% of the time on \texttt{Collision detection}. In molecular dynamics simulations, pairwise interactions similar to this are typically the dominating part. Here, however, the main part of the simulation was spent in the \texttt{Integrator} with over 61\%. The reason for this is a combination of the integrator's higher amount of floating-point operations per particle, as well as the fact that due to interface restrictions, all particles have to be traversed 10 times per iteration, leading to sub-optimal caching.
Overall, the simulation speed was about 0.01 seconds per iteration which advanced the simulation by 10 seconds or 10.91 hours for one year. With this, it was possible to simulate 20 years in 218.25 hours.

\section{Conclusion}
This work presents a proof of concept that deterministic conjunction tracking of space debris over a long time is feasible. Relying on state-of-the-art particle simulation methods combined with efficient choices for trajectory propagation and conjunction tracking, a large population was successfully modeled over 20 years. Thus, this model can serve to validate the well-established stochastic methods \cite{liou2004legend,lewis2012synergy,liou2003new} and bears the promise to provide a deterministic model of the future debris environment.
In the future, there are several natural continuations and extensions of this work. First, a breakup model will be incorporated to simulate the debris that is generated due to the found conjunctions to investigate the possible impact on critical orbits through exponential effects such as Kessler syndrome \cite{kessler1978collision}. Secondly, additional accuracy should be obtainable by employing a higher-order interpolation during the conjunction tracking (see \Cref{sec:conjunctionTracking}). Ideally, already computed positions and velocities at additional sub-timesteps from the integrator used for the propagation can serve to increase the interpolation order. A more sophisticated integrator targeted specifically at astrodynamics may improve results further \cite{blanes2013new}. Finally, the inclusion of small debris particles between 1 and 10~cm as well as modeling of planned upcoming mega-constellations can serve to study a more complex and realistic scenario.
\section*{Acknowledgments}

The authors would like to thank Stijn Lemmens and Tim Flohrer from ESA's Space Debris Office for their advise on this work.

\newpage

\bibliographystyle{model1-num-names}   
\bibliography{references}             

\vspace{0.25in}

%
%

\end{document}